\documentclass[12pt]{iopart}
\usepackage{iopams,epsfig}
\begin{document}

\title[Bosonization and entanglement spectrum for 1D disordered polar bosons]{Bosonization and entanglement spectrum for one-dimensional polar bosons on disordered lattices}

\author{Xiaolong Deng}
\address{Institut f\"ur Theoretische Physik, Leibniz Universit\"at Hannover, Appelstr. 2, D-30167 Hannover, Germany}
\ead{Xiaolong.Deng@itp.uni-hannover.de}
\author{Roberta Citro}
\address{Dipartimento di Fisica ``E. R. Caianiello'' and Spin-CNR, Universit\`a degli Studi di Salerno, Salerno, Italy}
\author{Edmond Orignac}
\address{Laboratoire de Physique de l'\'Ecole Normale Sup\'erieure de Lyon, CNRS-UMR5672, 69364 Lyon Cedex 7, France}
\author{Anna Minguzzi}
\address{Universit\'e Grenoble-Alpes and CNRS, Laboratoire de Physique et Mod\'elisation,
des Milieux Condens\'es UMR 5493, Maison des Magist\`eres, B.P. 166, 38042 Grenoble, France}
\author{Luis Santos}
\address{Institut f\"ur Theoretische Physik, Leibniz Universit\"at Hannover, Appelstr. 2, D-30167 Hannover, Germany}

\begin{abstract}

The extended Bose-Hubbard model subjected to a disordered potential  is predicted to display a rich phase diagram. In the case of uniform random disorder one finds  two insulating quantum phases -- the Mott-insulator and the Haldane insulator --  in addition to a superfluid and a Bose glass phase. In the case of a quasiperiodic potential further phases are found, eg the incommensurate density wave, adiabatically connected to the Haldane insulator. For the case of weak random disorder we determine the phase boundaries using a perturbative bosonization approach. We then calculate the entanglement spectrum for both types of disorder, showing that it provides a good indication of  the various phases.
\end{abstract}

\submitto{\NJP}
\maketitle

\section{Introduction}
The extraordinary experimental advances on the realization and control of ultracold quantum gases subjected to optical lattice potentials \cite{RMPBloch2008} pave the way  to the application of these systems as 'quantum simulators' \cite{BloDal2012}, capable of exploring with unprecedented accuracy complex models from condensed matter physics \cite{Bloch-FermiHubbard,Esslinger-FermiHubbard} to high-energy physics \cite{Bloch-higgs}.

The investigation of the interplay of disorder and interaction effects remains one open question in condensed matter physics, linked to the study of the metal-insulator transition. For bosonic systems, the Bose glass phase \cite{Giamarchi1988,Fisher1989} is an example of a novel strongly correlated phase arising from the simultanous effect of disorder and interactions. Most of solid-state based physical systems have to deal with some amount of disorder, which originates e.g. from defects in the material or impurities atoms.  A very peculiar feature of quantum gases is that one can add a tunable and controllable amount of disorder in  the pure system. In the regime of vanishing interactions, Anderson localization has been observed, first in one spatial dimension \cite{Roati2008,Billy2008} and then also in  three dimensions \cite{Kondov2011,Jendrzejewski2012}. The Bose glass phase has also been explored with bosons with short-range interactions on a lattice \cite{Fallani2007}.

The very recent advances on trapping and cooling ultracold molecules \cite{Ospelkaus2010} and atoms with a large dipole moment \cite{Griesmaier2005,Lu2011} prelude to the exploration with atomic quantum simulators of yet another set of systems, those with long range interactions. For the case of one-dimensional bosons on a lattice, the minimal model accounting for longer-range interactions is the extended Bose-Hubbard model (EBHM), which includes next-neighbour interactions among the bosons. In the case of a clean system, the EBHM displays already a rich phase diagram, which features novel several insulating phases, ie the density wave and the Haldane insulator, in addition to the Mott insulator already found in the Bose-Hubbard model \cite{Fisher1989}.

In this context, a relevant question arises on the fate of such insulating phases under the effect of disorder, as well on which novel phases arise in the presence of disorder. In the strong-coupling regime and for unitary lattice filling, a good starting point to understand this behaviour is to map the problem onto spin chains. For the latter system, the stability of the various insulating phases  at weak disorder has been studied  using renormalization group arguments \cite{Brunel1998}. For example, the density wave phase is found to disappear at infinitesimally small disorder, according to the Imry-Ma argument \cite{Imry1975,Shankar1990}. In analogy to the results known for the Bose-Hubbard model, other insulating phases are expected to shrink in the phase diagram, in favour of disordered correlated phases of the Bose-glass type.  The mapping onto spin systems fails when the tunnel energy becomes sufficiently important with respect to repulsive interactions to allow large occupancy of given lattice sites, and a superfluid, gapless phase builds up. In the general case then the problem has to be addressed numerically, and we have recently explored it using  Density-Matrix Renormalization Group \cite{Deng2012}, establishing a phase diagram by following  the behaviour of several correlation functions. In this work, we complement and support the phase diagram  using two complementary tools:  a bosonization and renormalization group description at weak disorder, and the study of the entanglement spectrum for the system.

\section{The model, numerical methods and phase diagrams}
We consider a system of N dipolar bosons confined onto  a one-dimensional deep optical lattice and in the presence of a very shallow trapping potential. We assume the dipole orientation to be  perpendicular to the lattice direction and truncate the dipole-dipole interaction potential to nearest-neighbour interactions.  Although longer range interactions do play a role, for very weak interactions and sufficiently large dipoles, the most relevant properties of polar physics can be already understood from a model of nearest neighbour interaction. This leads to the Hamiltonian of the extended Bose-Hubbard model (EBHM)
\begin{equation}
H = -t\sum_{i} (b^{\dagger}_i b_{i+1}+h.c.) + \frac{U}{2}\sum^N_{i=1} n_i(n_i -1) + V\sum_{i} n_in_{i+1} + \sum_i \epsilon_i n_i,
\end{equation}
where $b^{\dagger}_i$, $b_i$ are creation and annihilation operators for bosons at site $i$,  $n_i=b^{\dagger}_ib_i$ are the number operators,
$t$ is the tunnel energy, and  $U$~($V$) are the on-site~(nearest-neighbour, NN) interaction energies. Disorder is included in the model either through the random on-site energies  $\epsilon_i$, chosen to be uniformly distributed in the interval $[-\Delta, \Delta]$, or by a quasiperiodic potential $\epsilon_i=\Delta \cos(2 \pi \alpha i + \phi)$ and $\alpha$ is a rational number.

\subsection{Numerical method}

We have studied the EBHM using the DMRG method with open boundary conditions. The considered system sizes range up to 233 sites and we have taken up to 60 disorder realizations per point. In order to avoid the presence of metastable states we allow the number of optimal states to shrink or expand at every DMRG step according to a two-step algorithm which keeps at least one of the eigenvectors in the blocks of the reduced density matrix if they have only zero eigenvalues, and then keeps an additional eigenvector with zero eigenvalue in the block with non-zero eigenvalues if the eigenvalues decay very sharply to zero.  Furthermore, we have eliminated the edge states in the HI phase by adding one more particle or by coupling two extra hard-core bosons at the edges of the chain in order to form a singlet state~\cite{Deng2011}.

\subsection{Phase diagram of the EBHM without disorder}
The phase diagram of the EBHM in absence of disorder is known \cite{DallaTorre2006,Berg2008} and is illustrated in Fig.\ref{fig1}.
If the tunnel energy is dominating on the on-site interactions the bosons are delocalized throughout the lattice and the system is superfluid. The fluid displays quasi-long range order, with an algebraic decay of the first-order correlation function  $G(r)=\langle b_i^\dag b_{i+r}\rangle /\sqrt{\langle n_i\rangle\langle n_{i+r} \rangle}\propto r^{-1/2K}$. At increasing on-site interactions and weak NN interactions a Kosterlitz-Thouless transition occurs towards the incompressible Mott-insulator (MI) phase, characterized by a hidden parity order ${\cal O}_{P}=\lim_{|i-j|\rightarrow\infty}\langle (-1)^{\sum_{i<l<j}\delta n_l}\rangle$~\cite{Berg2008}, , where $\delta n_l=1-n_l$. At sufficiently large values of $U$ and intermediate values of $V$ a second insulating phase is found. This Haldane insulator (HI) is characterized by a hidden string order ${\cal O}_{S}=\lim_{|i-j|\rightarrow\infty}\langle \delta n_i (-1)^{\sum_{i<l<j}\delta n_l}\delta n_j\rangle$. At sufficiently large values of $V$ and $U$ a third insulating phase occurs, a density wave with spatial modulation (DW) as can be identified by the finite correlator  ${\cal O}_{DW}=\lim_{|i-j|\rightarrow\infty}\langle (-1)^{i-j}\delta n_i \delta n_j\rangle$. At sufficiently large $U$ a direct first-order transition is expected between  Mott-insulator and density-wave.

\subsection{Phase diagram of the disordered EBHM}
\begin{figure}
\includegraphics[width=0.3\linewidth]{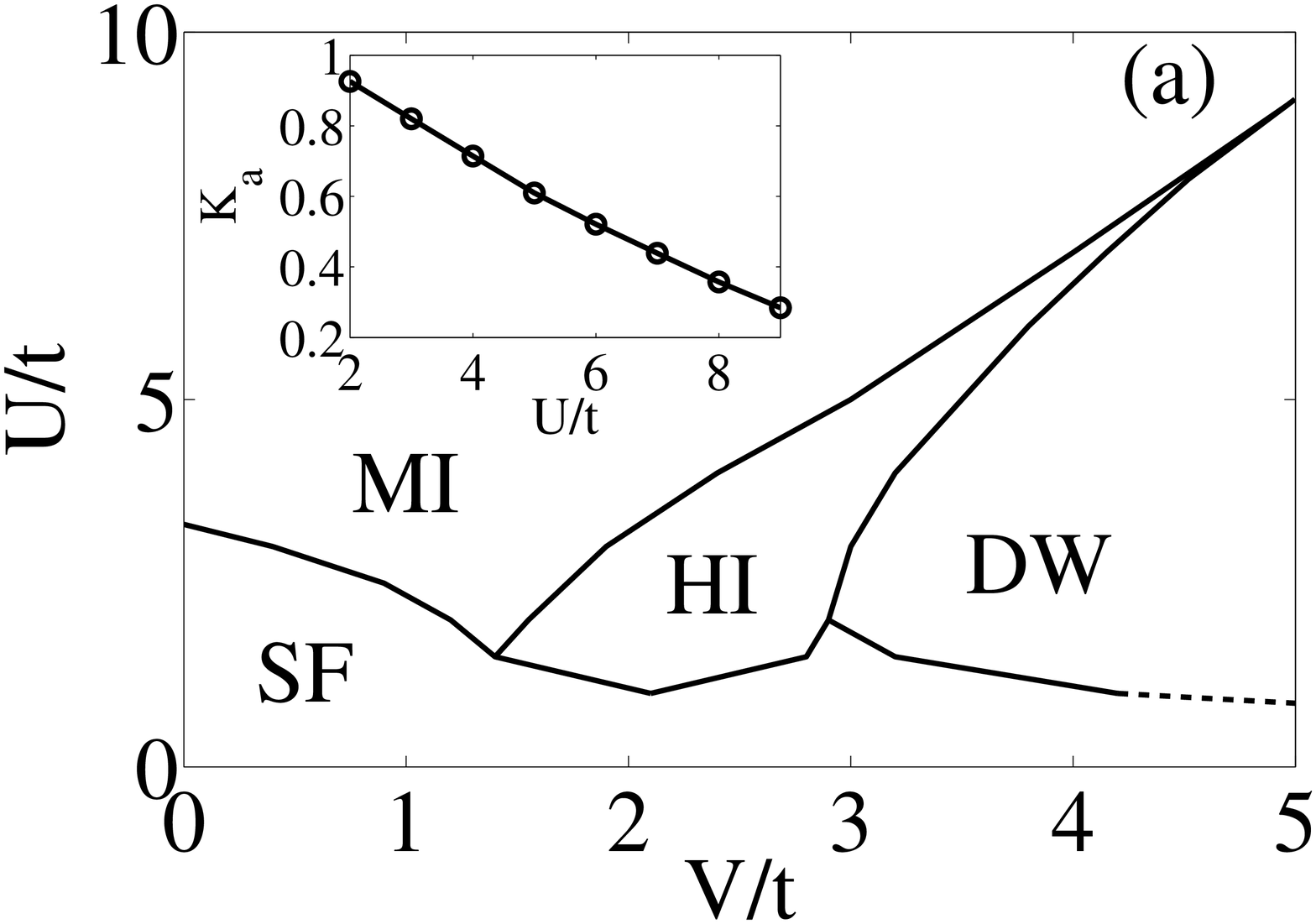}
\includegraphics[width=0.3\linewidth]{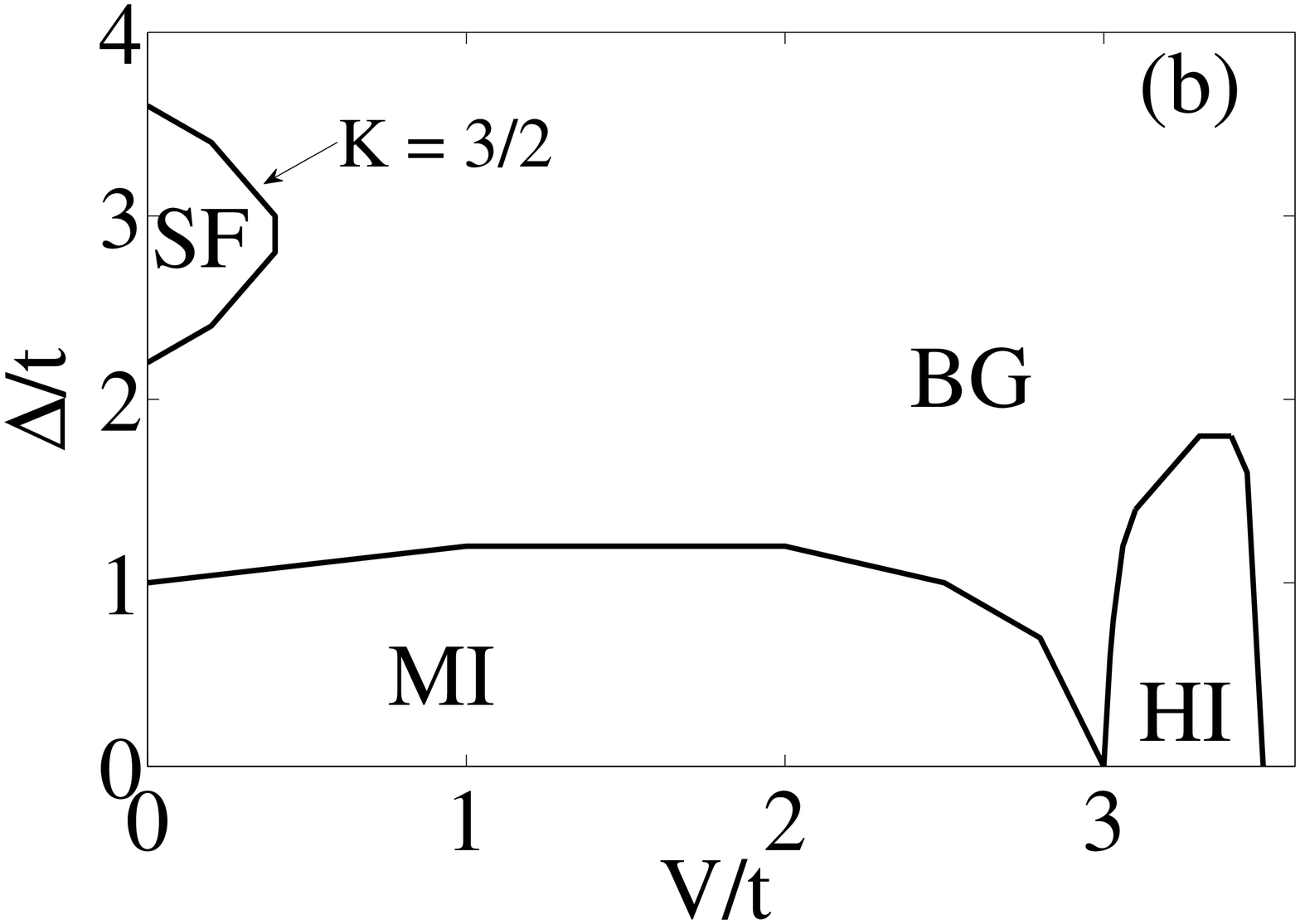} \includegraphics[width=0.3\linewidth]{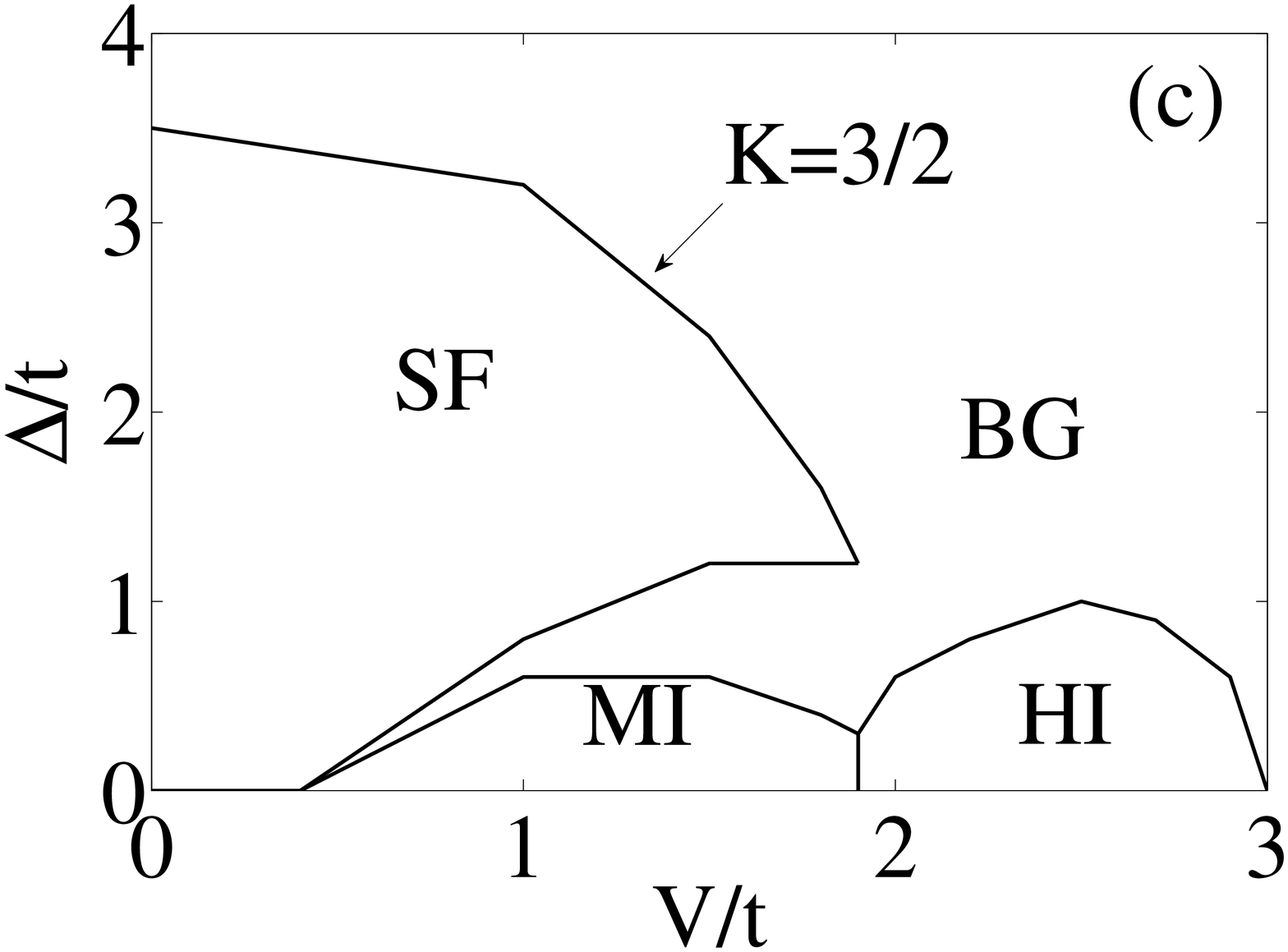}
\caption{Phase diagram for  polar bosons with nearest-neighbor interactions. Left panel:  clean case  in the plane $(V/t,U/t)$, with inset showing the value of the Luttinger parameter $K_a$ along the MI-HI critical line, which is described by a Luttinger model. The HI-DW critical line belongs to the Ising universality class \cite{DegliEsposti}. Central and right panel:  case of uniform disorder,  in the plane $(V/t,\Delta/t)$ for $U/t=5$ and $U/t=3$.}
\label{fig1}
\end{figure}

The phase diagram of the EBHM in presence of uniform disorder has been first studied in \cite{Deng2012} and is illustrated in Fig.\ref{fig1}. Among the phases of the pure system, the density wave phase is unstable under infinitesimal disorder according to the Imry-Ma argument \cite{Imry1975} thus disappears. Mott insulator and Haldane insulator disappear at sufficiently strong disorder to leave place to a compressible, non-superfluid Bose-glass phase~\cite{Giamarchi1987,Giamarchi1988,Fisher1989}. An additional superfluid lobe is found at finite disorder as in found in Ref.~\cite{Rapsch1999,Prokofev1998} for $V=0$, and corresponds to the regime where repulsions overcome localization effects~\cite{Giamarchi1988}.

 It is not easy to resolve numerically  the behavior of the critical MI-HI point for weak disorder, and in  particular whether the critical line remains stable or whether a Bose glass  intermediate region opens at $\Delta=0$. This issue is discussed in Sec.~\ref{sec:bosonization} using bosonization and renormalization (RG) arguments.

\subsection{Phase diagram of the EBHM with quasiperiodic potential}
For the case of a quasiperiodic potential the phase diagram,  shown in Fig.\ref{fig2},  considerably differs from the one with uniform disorder \cite{Deng2012}. The main features are the presence of an incompressible density wave phase, typical of the quasiperiodic potentials \cite{Roscilde2008,Roux2008}, adiabatically connected to the Haldane insulator phase, and the persistence of a density wave phase, due to the intrinsecally different nature of the quasiperiodic potential with respect to a truly random potential \cite{Leboeuf}.

\section{Bosonization approach at weak disorder}
\label{sec:bosonization}

For sufficiently strong interactions, where number fluctuations on
each site are relatively small, we truncate the occupancy of each site
to the values $\{0,1,2\}$. We then employ the Holstein-Primakoff
transformation $S_i^z=\delta n_i=1-n_i$, $S_i^+=\sqrt{2-n_i}b_i$. to
map the EBHM onto a spin-1 Hamiltonian with single-ion anisotropy
\begin{equation}
 H=-2 t\sum_i S_i^xS_{i+1}^x+S_i^yS_{i+1}^y+V \sum_i S_i^zS_{i+1}^z+ U/2 \sum_i (S_i^z)^2 \label{eq:H-Spin}
\end{equation}
Following the early works of Timonen and Luther and Schulz
\cite{TimLut85,Sch86} we represent the spin-1 operators as the sum of
two spin 1/2 operators,
$S_i^\alpha=\sigma_{i,1}^\alpha+\sigma_{i,2}^\alpha$. This brings the
Hamiltonian (\ref{eq:H-Spin}) into the one of two coupled spin
chains. Furthermore, a Jordan-Wigner transformation is employed to map
the spin-1/2 operators onto fermions according to
$\sigma_{i,\lambda}^z=a^\dagger_{i,\lambda}a_{i\lambda}-1/2 $,
$\sigma_{i,\lambda}^+=a^\dagger_{i,\lambda} e^{i \pi
  \sum_{n=1}^{i-1}} a^\dagger_{n,\lambda} a_{n,\lambda}$, with
$\lambda=1,2$. A continuum limit $a_{n,\lambda}=\sqrt{a}
\sum_{p=\pm} \psi^\lambda_p(n a)$ where $a$ is the lattice spacing and
$p$ stands for $\pm$, left or right mover  is taken. 
Finally,
one employs a low-energy description of each fermionic field,
$\psi^{\lambda}_p(x)\sim\frac{1}{2\pi\alpha} e^{i p k_F x} e^{-i(p
  \phi_{\lambda}(x)-\theta_{\lambda}(x))}$, where the fields
$\theta_{\lambda}(x)$ and $\phi_{\lambda}(x)$ satisfy canonical
conjugation relations $[ \phi_{\lambda}(x),\partial_x
\theta_{\lambda'}(x')]=i \pi \delta_{\lambda \lambda'}\delta(x-x')$.
We have $k_F=\pi/(2a)$ when $\langle S^z \rangle=0$ i. e. the filling
is one boson per site.     
  This leads to the Hamiltonionan of two coupled
Tomonaga-Luttinger fluids \cite{Sch86,Orignac1998,Brunel1998}, 
which takes a simple form $H=H_a+H_o$ once the
'acoustical' and 'optical' combinations are introduced
$\phi_a=(\phi_1+\phi_2)/\sqrt{2}$, $\phi_o=(\phi_1-\phi_2)/\sqrt{2}$,
and similarly for the $\theta_\lambda$ fieds, 
\begin{equation}
\label{Eq:Ha}
H_a=\frac{\hbar u_a}{2\pi}\int dx \lbrack K_a (\partial_x
\theta_a)^2+\frac{1}{K_a} (\partial_x \phi_a)^2 \rbrack+
\frac{g_1}{(\pi a)^2} \int dx \cos(\sqrt{8} \phi_a) 
\end{equation}
\begin{eqnarray}
\label{Eq:Ho}
H_o&=&\frac{\hbar u_o}{2\pi}\int dx \lbrack K_o (\partial_x
\theta_o)^2+\frac{1}{K_o} (\partial_x \phi_o)^2\rbrack
+\frac{g_2}{(\pi a)^2} \int dx \cos(\sqrt{8} \phi_o)\nonumber \\
&+&\frac{g_3}{(\pi \alpha)^2} \int dx \cos(\sqrt{2} \theta_o) 
\end{eqnarray}
where $g_1=g_2=(U-V)a$, $g_3=-2t a$.  Coupling between acoustical and
optical sectors is found at higher order\cite{DallaTorre2006,Berg2008} 
and is therefore less
relevant than the terms listed here. The weak-coupling expressions for
the Luttinger parameters entering Eqs.(\ref{Eq:Ha}) and (\ref{Eq:Ho})
read
\begin{eqnarray}
u_a=2 t a \sqrt{1+ \frac{U+6V}{2 \pi t a}} , \mbox{ } K_a=\frac{1}{\sqrt{1+ \frac{U+6V}{2 \pi t a}} }\\
u_o=2 t a \sqrt{1- \frac{U-V}{2 \pi t a}} , \mbox{  } K_o=\frac{1}{\sqrt{1-\frac{U-V}{2 \pi t a}} }
\end{eqnarray}
Introducing the rescaled fields $\theta_{+/-}=\theta_{a/o}/\sqrt{2}$ and
$\phi_{+/-}=\sqrt{2} \phi_{a/o}$, the Hamiltonians in
Eqs.~(\ref{Eq:Ha}) and (\ref{Eq:Ho}) can be brought to the form used
in Refs.~\cite{DallaTorre2006,Berg2008}. 

The phase diagram \cite{Sch86} can be deduced from (\ref{Eq:Ha}) and
(\ref{Eq:Ho}). For $K_a>1$ in (\ref{Eq:Ha}), the cosine term is
irrelevant, and the 'acoustic' modes are gapless. For $K_a<1$ the
cosine term is relevant, and the field $\phi_a$ is pinned either to $0$ for
$g_1<0$ or to $\pi/\sqrt{8}$ for $g_1>0$, the spectrum of $H_a$ being
always gapped. Meanwhile, in
(\ref{Eq:Ho}), at least one of the two cosines is relevant so the
spectrum is always gapful. Depending on the parameters, either
$\theta_o$ or $\phi_o$ is pinned. When $\phi_o$ (resp. $\theta_o$) 
is pinned, the
correlation functions of its dual fields $e^{i \beta \theta_o}$
(resp. $e^{i\beta \phi_o}$) decay
exponentially. Combining the different possibilities, we obtain the
two gapless superfluid phases (when $K_a>1$ and either $\phi_o$ or
$\theta_o$ is pinned), the gapful density wave phase (when both
$\phi_a$ and $\phi_o$ are pinned), and two phases with $\phi_a$ pinned 
and $\theta_o$ pinned. The phase with $\phi_a$ pinned to zero (i. e. $g_1<0$) is the
Mott insulator, while the phase with $\phi_a$ pinned to $\pi/\sqrt{8}$
(i. e. $g_1>0$) 
is the Haldane insulator \cite{DallaTorre2006,Berg2008}.   

We focus in the following on the parameter regime corresponding to the
Haldane-insulator to Mott-insulator transition point at weak
disorder. This regime is difficult to access numerically, but is
amenable to a perturbative renormalization group calculation.  
At the critical point $g_1=0$ which separates the Mott-insulating
from the Haldane insulating phase, the system is in a Luttinger liquid
state \cite{Chen2003,EspostiBoschi2003,Albuquerque2009,Hu2011} and the boson Green's function decays as
\begin{equation}
\langle b^{\dagger}(x) b(0)\rangle=|C|^2\left( \frac{\alpha}{|x|}\right)^{\frac{1}{4 K_a}}.
\end{equation}


One kind of randomness that we are able to treat is the on-site
disorder of the form $\sum_n \epsilon_n b^{\dagger}_n b_n$, which for the spin-1
chain corresponds to the effect of a random field along the
$z$-axis \cite{Orignac1998,Brunel1998}. 
When dealing with the coupling to disorder, one has to bear
in mind that the bosonized expression of the boson field is a series
which contains higher order harmonics\cite{Haldane1981} 
 in the field $\phi_{1,2}$ from
which the boson number operator can be written as
\begin{equation}
b^\dagger_nb_n=\sum_{m=1}^{\infty} A_m \cos(m \sqrt{2} \phi_a-2m k_F
x)\cos(m \sqrt{2}\phi_o), 
\label{eq:deff}
\end{equation}
where $x=n a$. 
The term of order one has been treated in \cite{Brunel1998} and it is
relevant when $K_a+K_o<3$, while higher order terms have been
neglected. When the second order term is taken into account, $A_2
\cos(2\sqrt{2}\phi_a -4 k_F x)\cos(2\sqrt{2}\phi_o)$, and performing a
perturbation theory in $\cos(2\sqrt{2}\phi_o)$ at first order, an
effective coupling to disorder is generated of the form:
\begin{equation}\label{eq:4kf-disorder} 
H^z_{eff}=(\xi_{4k_F}(x) A_2 e^{i \sqrt{8}\phi_a}+h.c.),
\end{equation}
where $\xi_{4k_F}(x)$ is a gaussian random variable with
$\overline{\xi_{4k_F}(x)\xi^\star_{4k_F}(x')}=D_{eff}\delta(x-x')$. A
renormalization group treatment of such a term gives 
\begin{equation}
\frac{dD_{eff}}{dl}=(3-4K_a) D_{eff},
\end{equation}
and $D_{eff}$ is relevant when $K_a<3/4$. The terms with $m>2$ become
relevant for lower values of $K_a<3/m^2$. Thus, along the critical
line $g_1^\star=0$ one expects a stable Luttinger liquid line till
$K_a^\star=3/4$, and below it  a Bose glass phase  between the
Mott-insulating and Haldane-insulating phase takes place. \\
Let us note that even in a model where only the $m=1$ term is kept in the
expansion (\ref{eq:deff}), the term (\ref{eq:4kf-disorder}) will be
generated \cite{Orignac1998}  by 
integrating over the fluctuations of the field $\theta_o$. So the
condition of stability $K^\star>3/4$ is a generic one.  As a consequence of this analysis, we expect that the HI-MI transition point in the presence of disorder has a 'Y' shape for $K_a>3/4$ and a 'V' shape for $K_a<3/4$. This is in agreement with the numerical results shown in Fig.\ref{fig1}, where for $U/t=5$ with $K_a=0.6$ we have a 'V' case and for $U/t=3$ with $K_a=0.8$ we have a 'Y' case.

\section{Entanglement spectra}
The calculation of the entanglement spectrum is  a novel approach to identify the quantum phases in several models. It is particularly useful to bring insight into  topological phases and phases with nonlocal order parameter (see eg \cite{Li08,Calabrese08,Regnault09,Lauchli10}) and has been specifically analyzed for the case of the Bose-Hubbard model \cite{Deng2011,Alba12,Alba12b}.
The entanglement spectrum is defined as the spectrum $- \ln \lambda_i(L_A)$ of the effective Hamiltonian $ -\ln \rho_A$, obtained by partitioning the system density matrix into two parts A and B (of length $L_A+L_B=L$) and tracing over the B part.  It has been shown that the behaviour of the eigenvalues $\lambda_i(L_A)$ and their degeneracy differs in the various phases of the clean EBHM \cite{Deng2011}, thus allowing to infer the structure of the phase diagram. We show here how the study of the entanglement spectrum can also give useful information in the disordered and aperiodic case.

\begin{figure}
\includegraphics[width=0.5\linewidth]{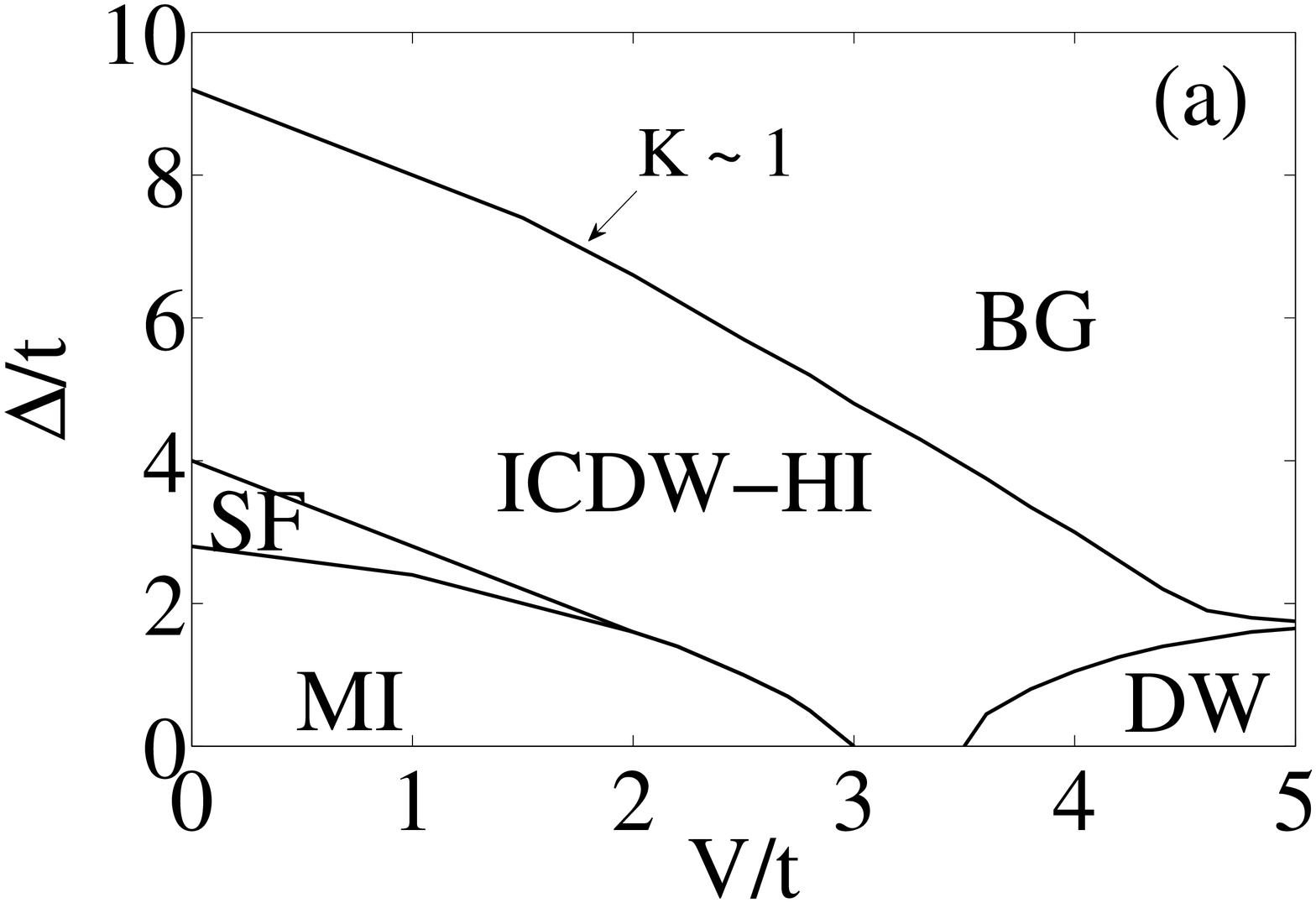}
\includegraphics[width=0.5\linewidth]{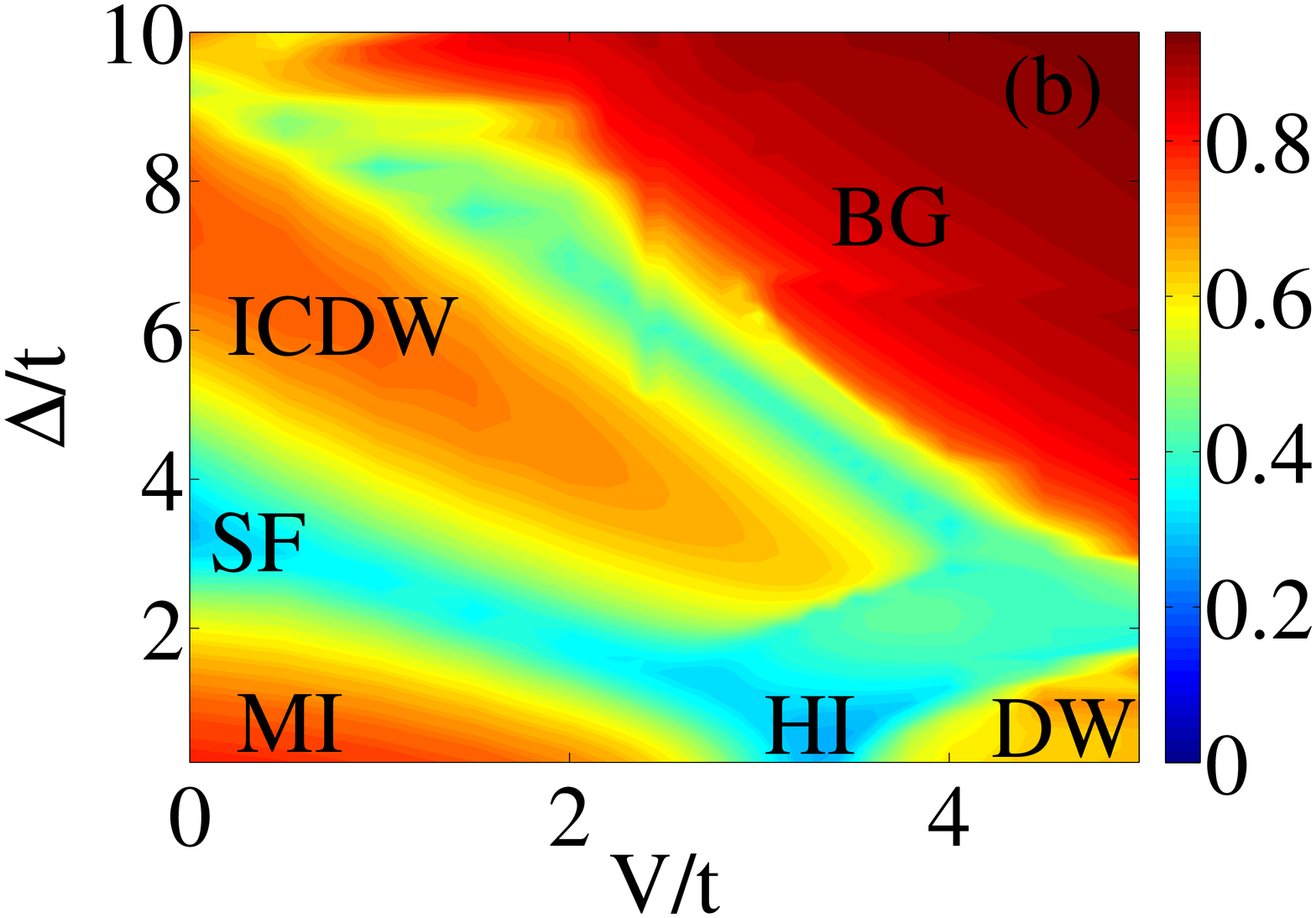}
\caption{Phase diagram for  polar bosons with nearest-neighbor interactions in a quasiperiodic potential  in the plane $(V/t,\Delta/t)$ for $U/t=5$. Left panel: phase diagram from the study of correlations functions as in \cite{Deng2012}. Right panel: results from the entanglement spectrum.}
\label{fig2}
\end{figure}

\begin{figure}
\begin{center}
\includegraphics[width=0.3\linewidth]{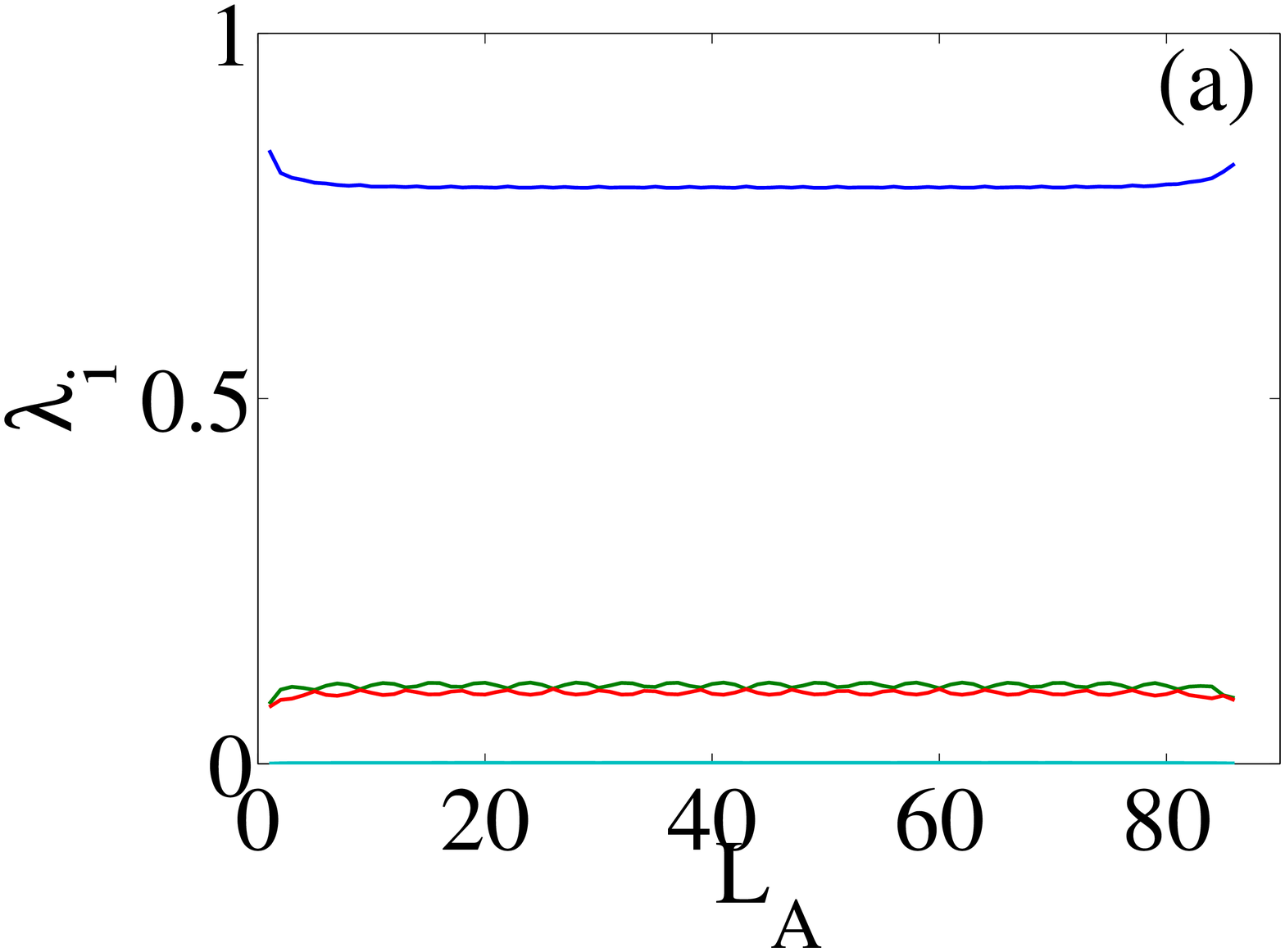}\includegraphics[width=0.3\linewidth]{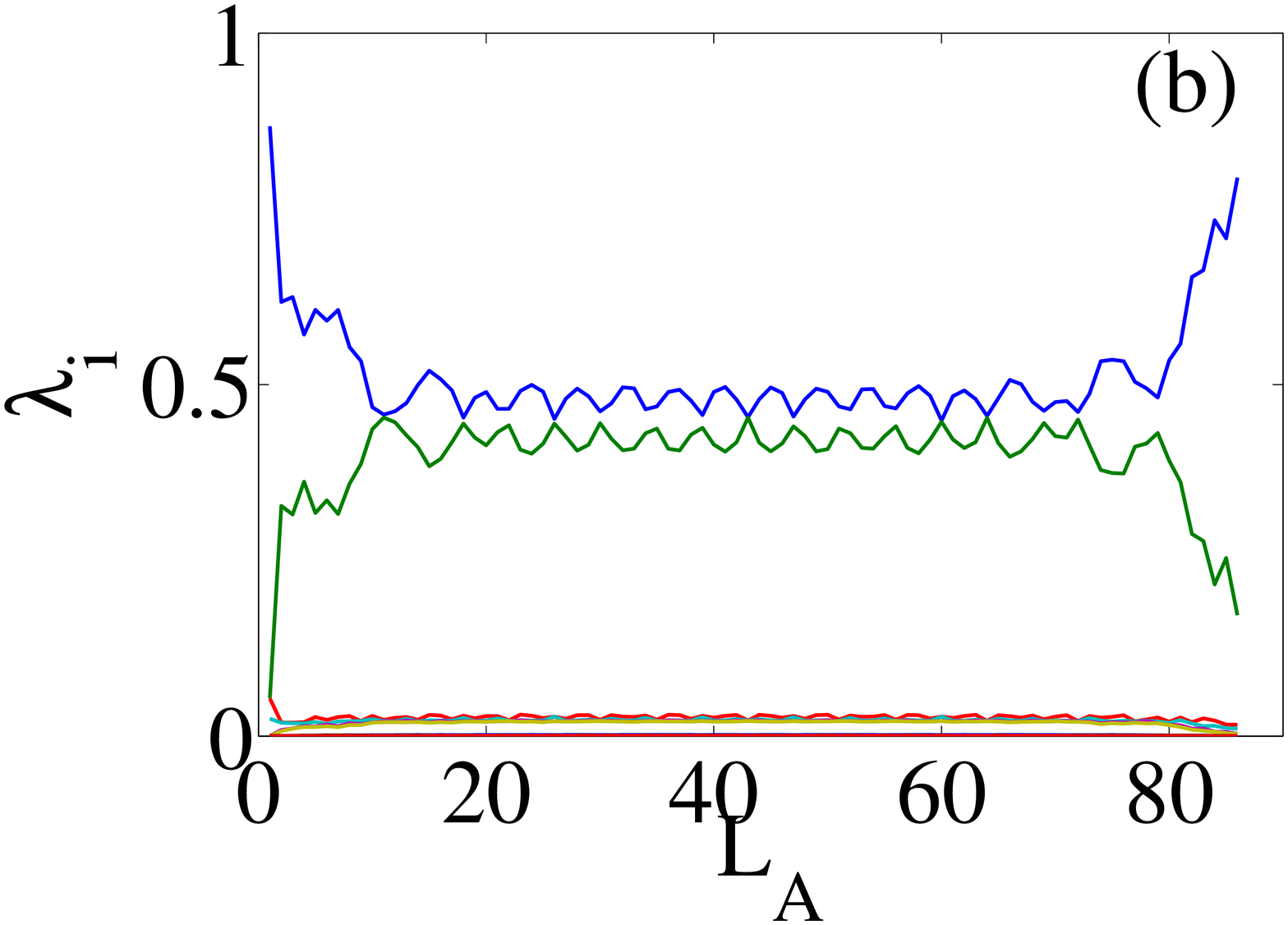}\includegraphics[width=0.3\linewidth]{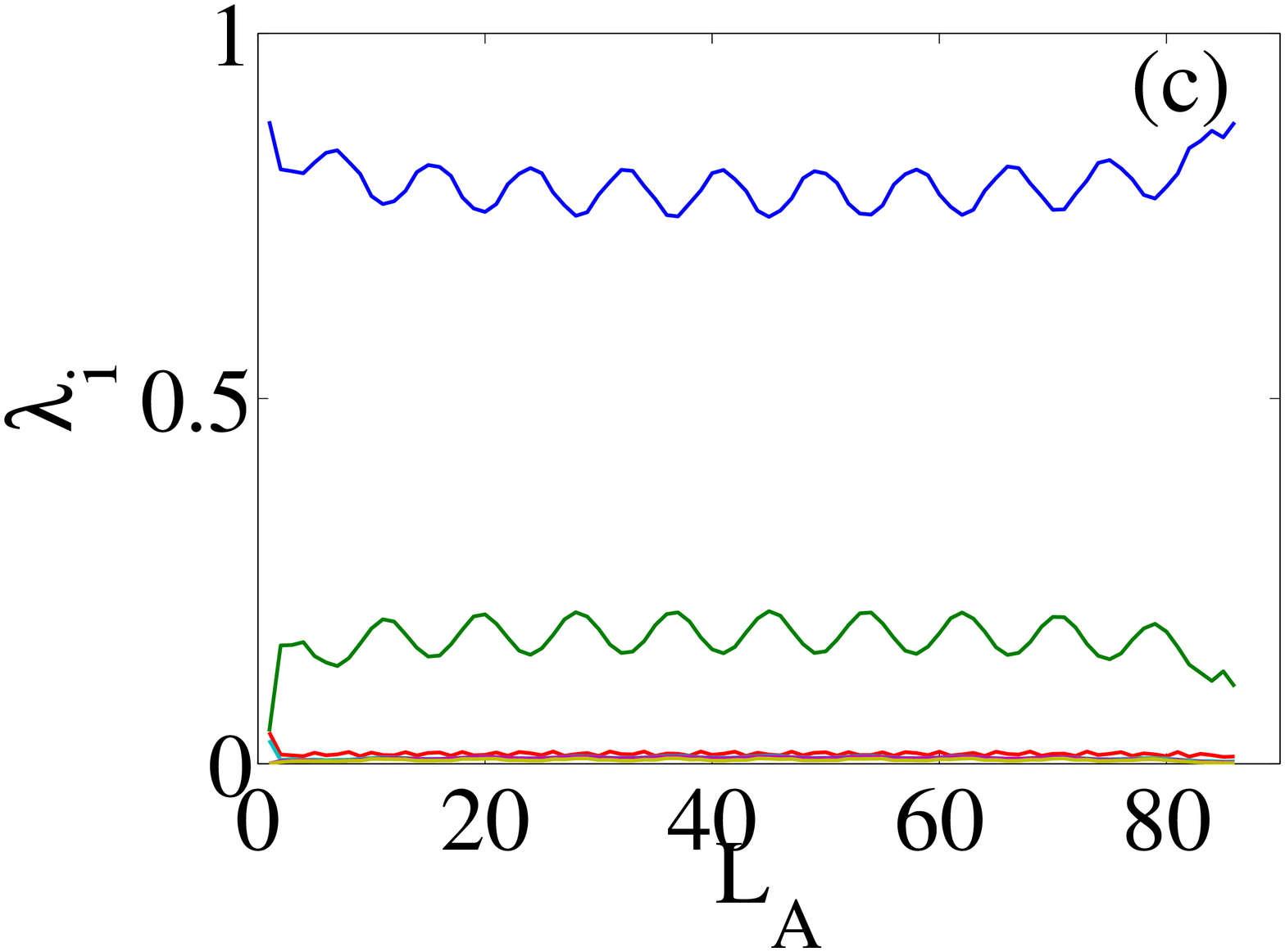}
\includegraphics[width=0.3\linewidth]{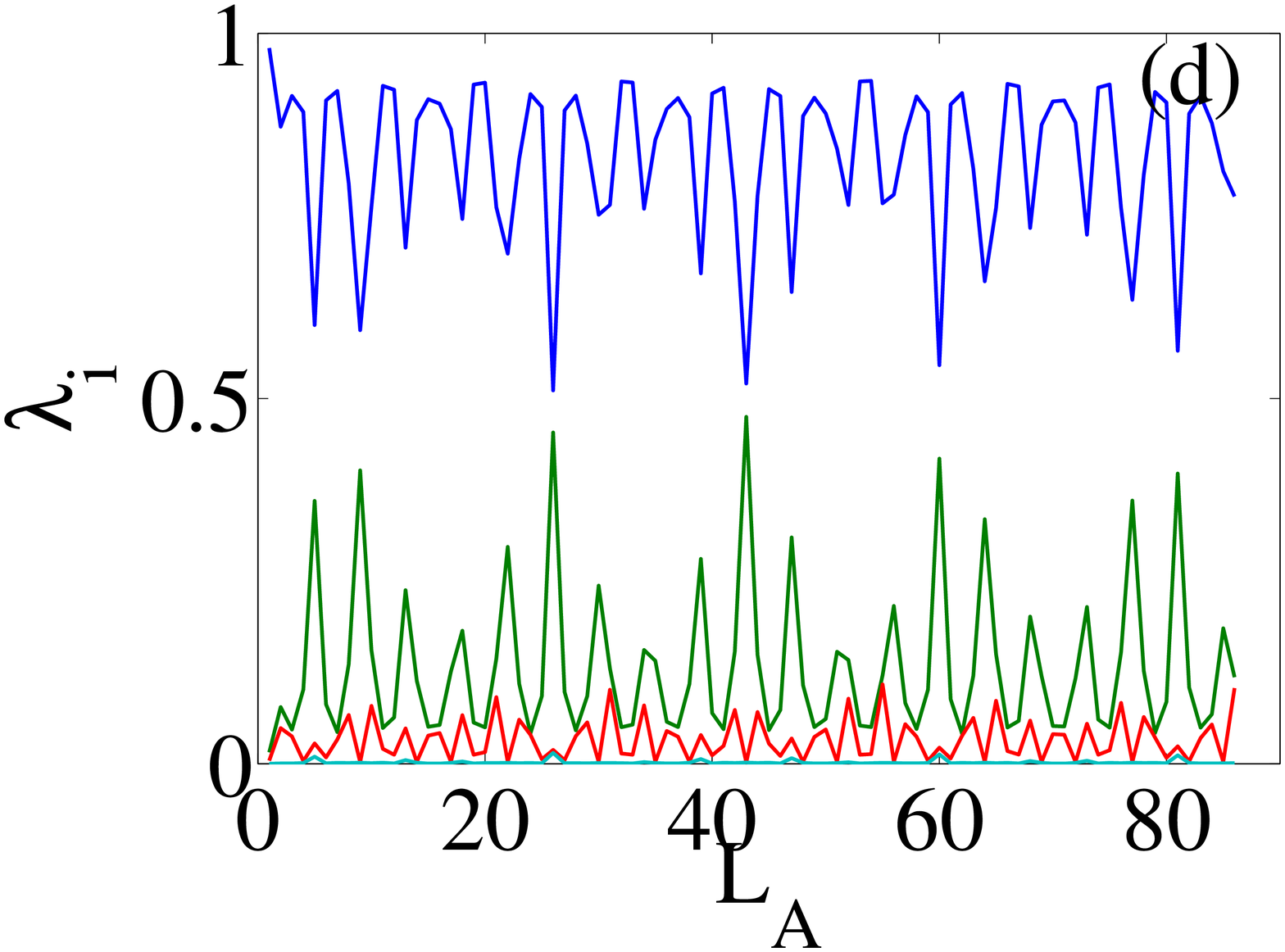}\includegraphics[width=0.3\linewidth]{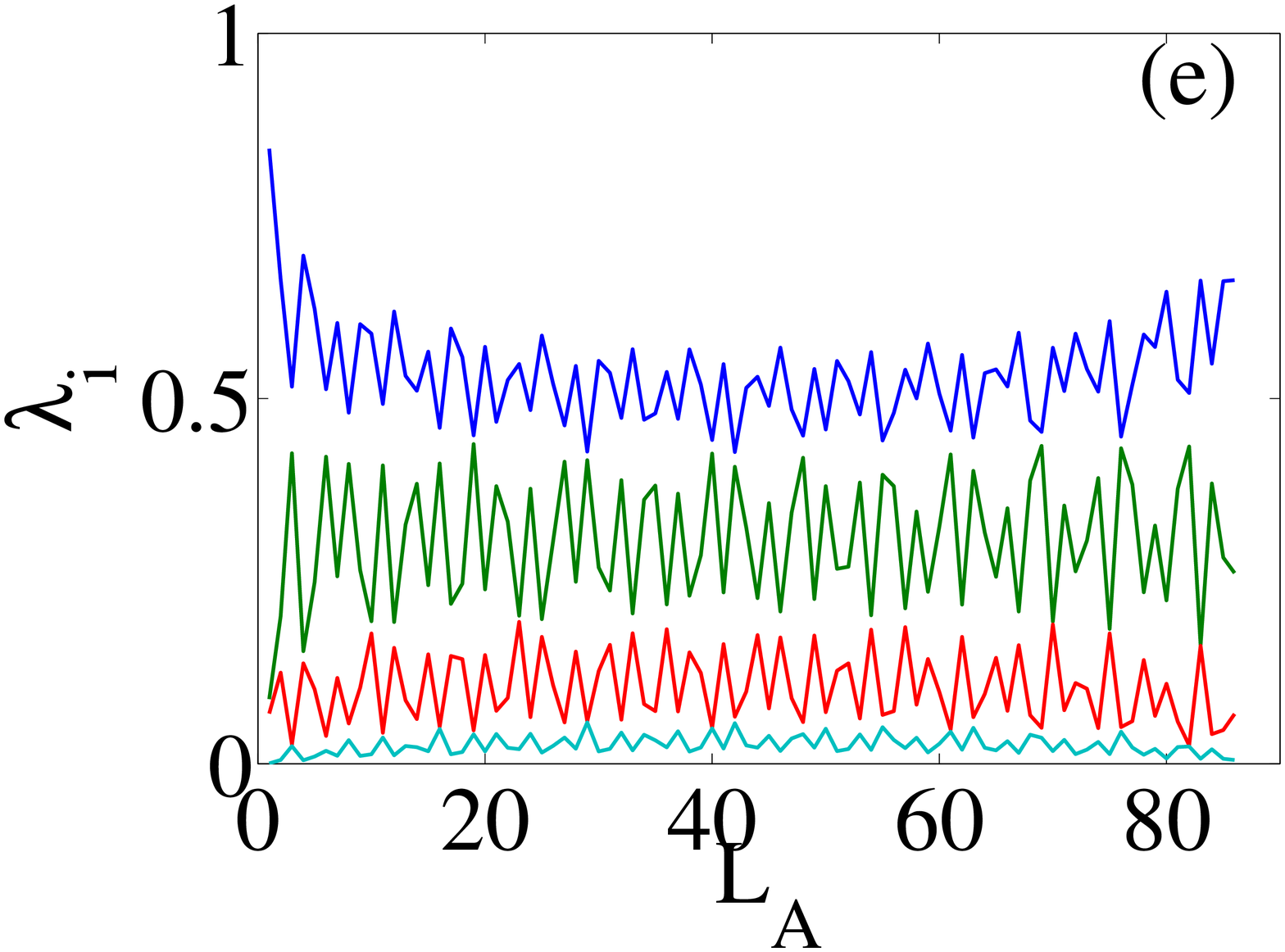}\includegraphics[width=0.3\linewidth]{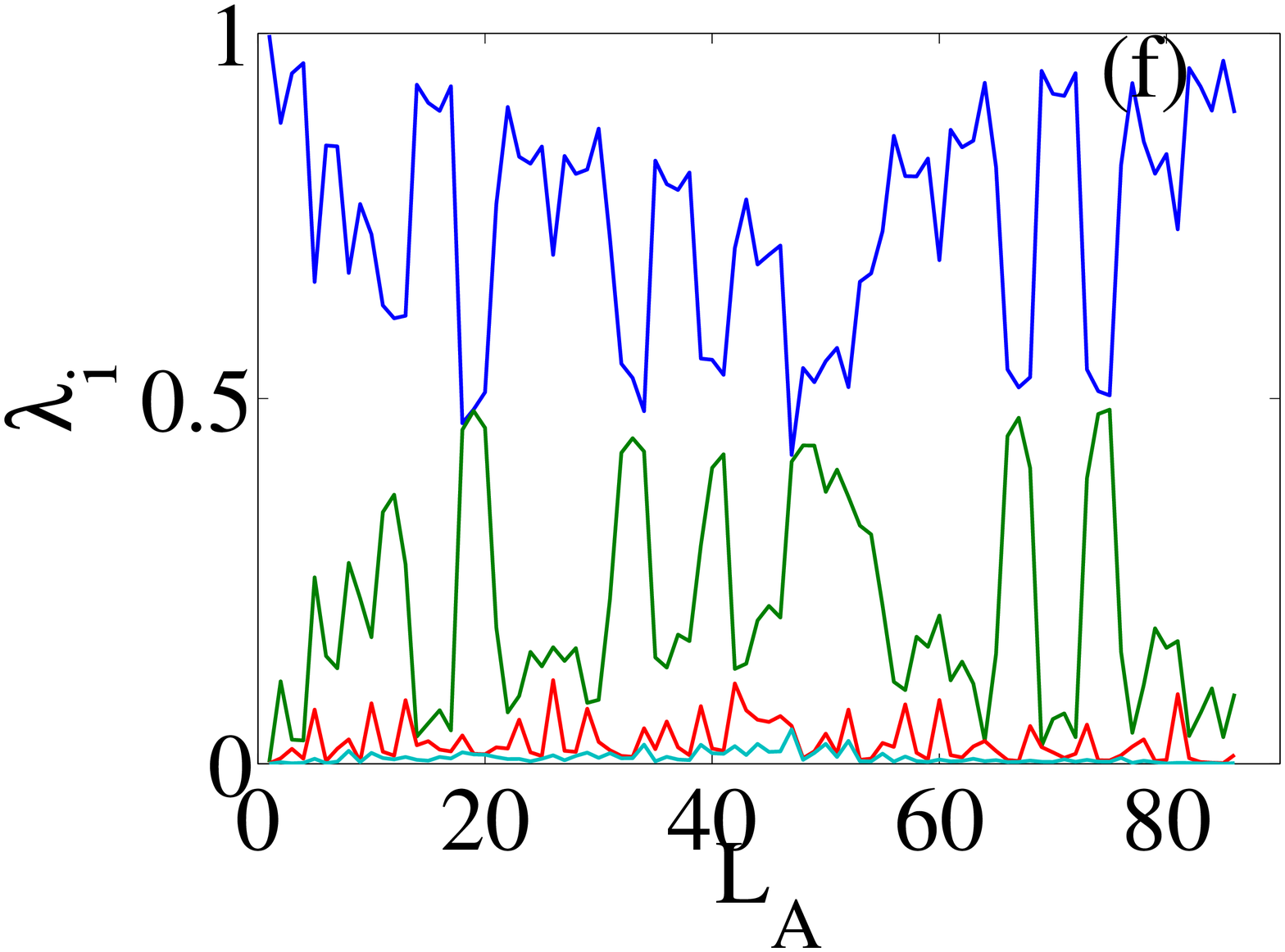}
\end{center}
\caption{Entanglement spectrum of the EBHM under quasiperiodic potential: largest eigenvalues as a function of the partition along the chain corresponding to a) MI at  $V/t=0.5$ $\Delta/t=0.2$; b) HI, at $V/t=3.3$,  $\Delta/t=0.2$; c) DW, at $V/t=3.6$, $\Delta/t=0.2$; d) ICDW, at $V/t=1$ $\Delta/t=6$; e) SF, at $V/t=0.5$, $\Delta/t=3$; f) BG, at  $V/t=2$, $\Delta/t=7.8$.}
\label{fig3}
\end{figure}

In order to obtain the phase diagram, we take the combination of the first four largest eigenvalues 
$\zeta=\lambda_1^T -\lambda_2^T + \lambda_3^T - \lambda_4^T$, where $\lambda_i^T=(1/L) \sum_{L_A=1}^L\lambda_i(L_A)$.  The result is illustrated in Fig.\ref{fig2}.  The alternating regions of low and large values of $\zeta$ have a very good correspondence to the phases predicted from the study of correlation functions. This is readily explained from the study of the behaviour of the largest eigenvalues $\lambda_i(L_A)$ in the various phases, of which some examples are given in Fig.\ref{fig3}. In the Mott insulator phase in the clean case a degeneracy is found between $\lambda_2^T$ and $\lambda_3^T$, as well as between  $\lambda_4^T$ and $\lambda_5^T$ while $\lambda_1^T$ is non degenerate. This feature is found to persist also in the quasiperiodic case, and yields a large value for $\zeta$. In the superfluid phase instead, the largest eigenvalues are almost equidistant; hence yielding a small $\zeta$. In the clean case, the Haldane insulator phase is characterized by a double degeneracy of the largest eigenvalues, hence implying another region of vanishing $\zeta$ \footnote{Since the simulation in Fig.\ref{fig2}(b) is done at integer filling the edge states cannot be eliminated,  hence a small breaking of the double degeneracy is found at vanishing disorder, as in \cite{Deng2012}.}. Such degeneracy is gradually broken at increasing the strength of the quasiperiodic potential. In the density wave phase no degeneracy is found in the clean case, thus allowing to clearly distinguish this phase from the neighbouring Haldane insulator; in the presence of a quasiperiodic potential, also this phase gradually disappears. A similar level structure is found for the incommensurate density wave. Finally, the BG phase displays a structure close to the  ICDW phase, with a few large non-degenerate eigenvalues. 


\begin{figure}
\begin{center}
\includegraphics[width=0.4\linewidth]{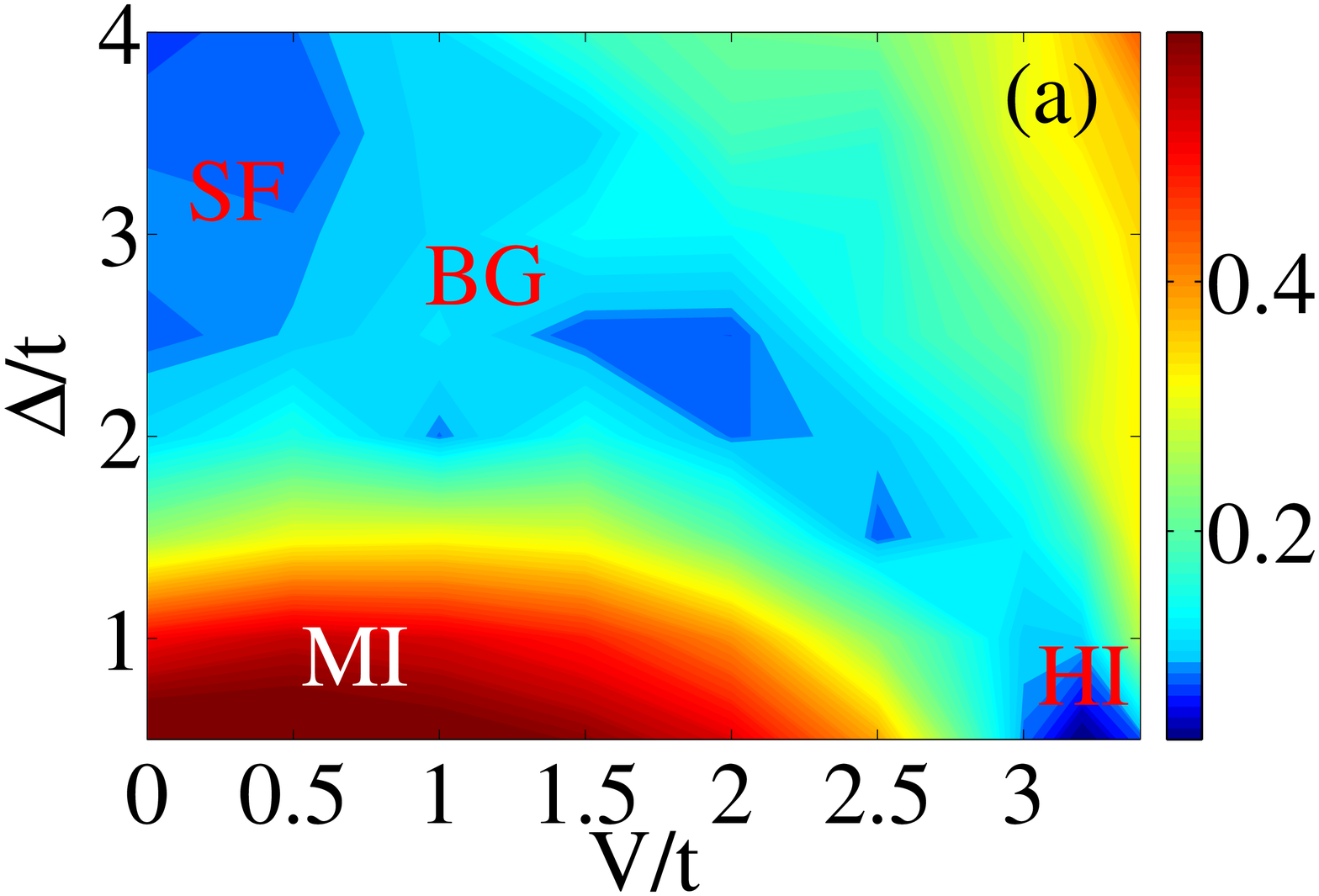}\includegraphics[width=0.4\linewidth]{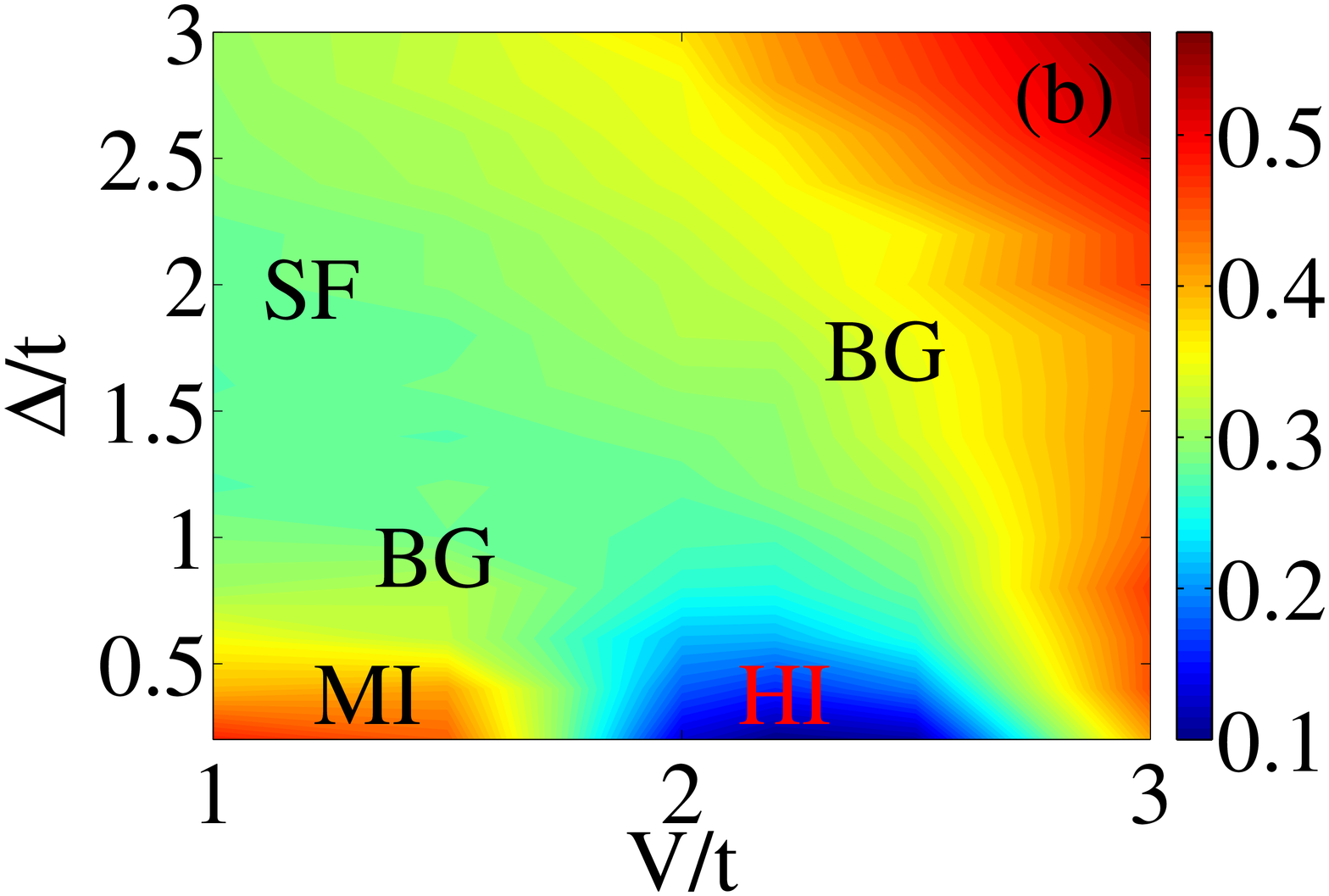}
\end{center}
\caption{Phase diagram of the uniform disorder case from the entanglement spectrum, (a) at $U/t=5$, (b) at $U/t=3$. Calculations performed with $L=55$ and averaging over 20 disorder realizations.}
\label{fig4}
\end{figure}

\begin{figure}
\begin{center}
\includegraphics[width=0.3\linewidth]{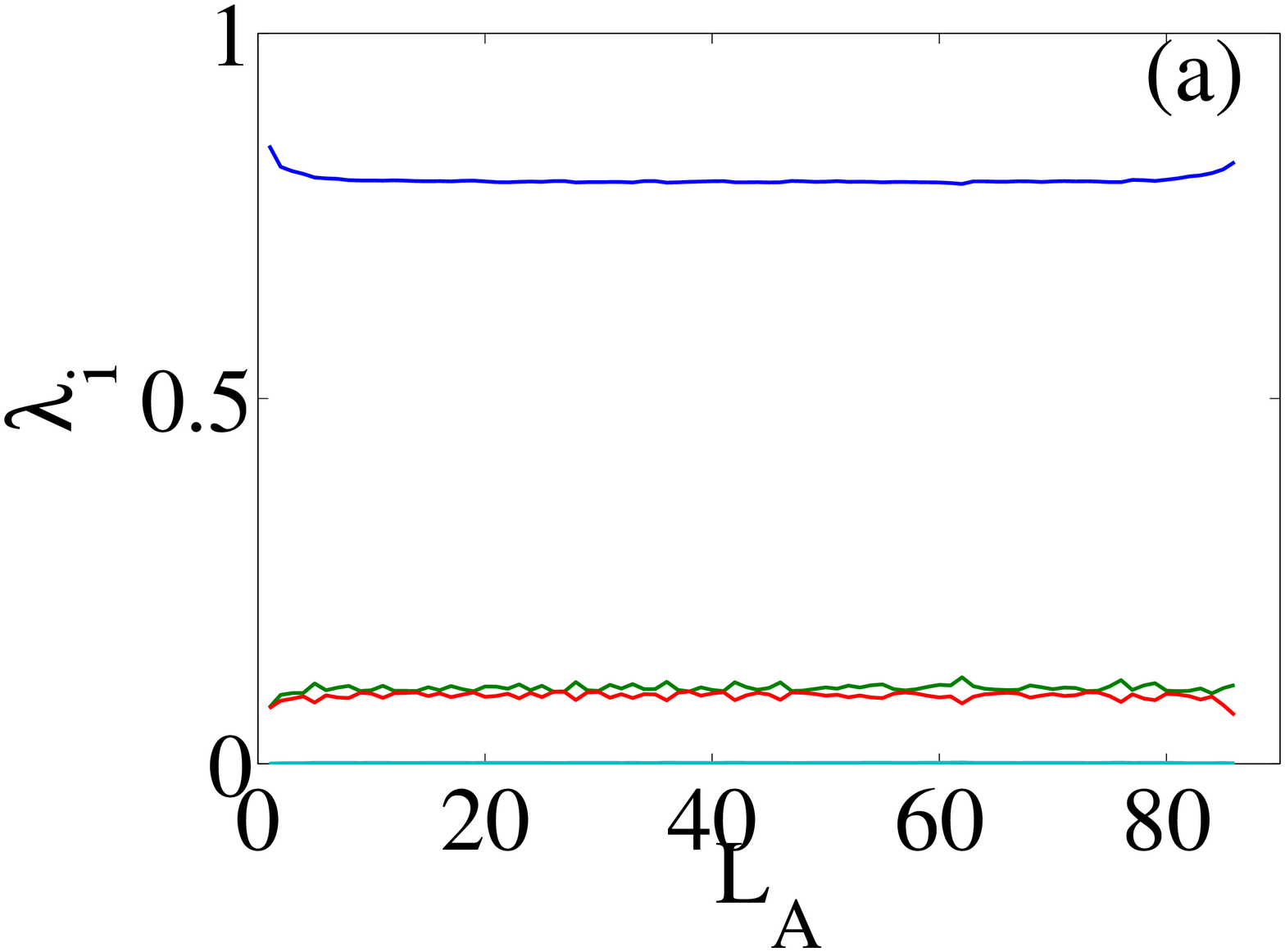}\includegraphics[width=0.3\linewidth]{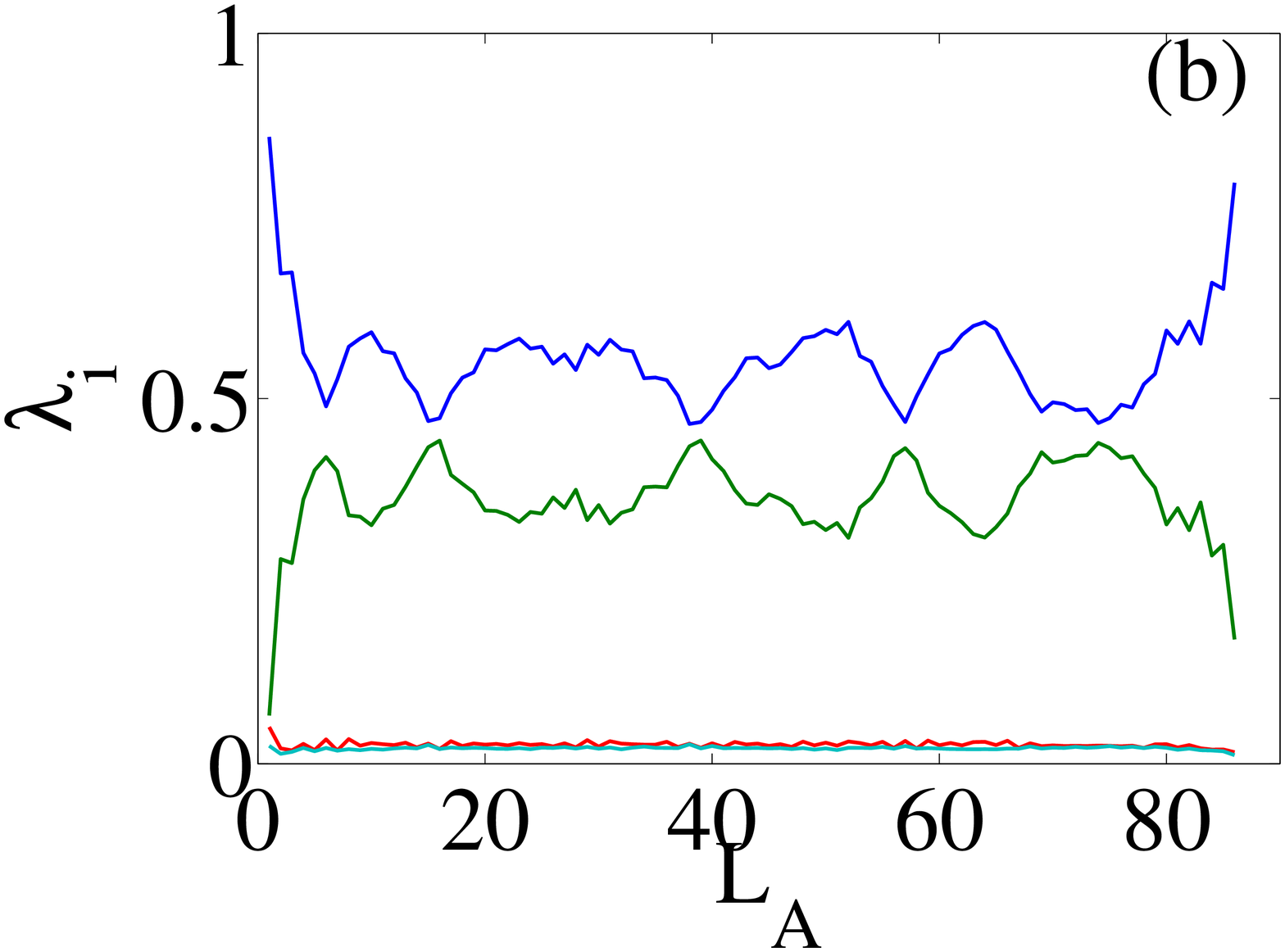}\includegraphics[width=0.3\linewidth]{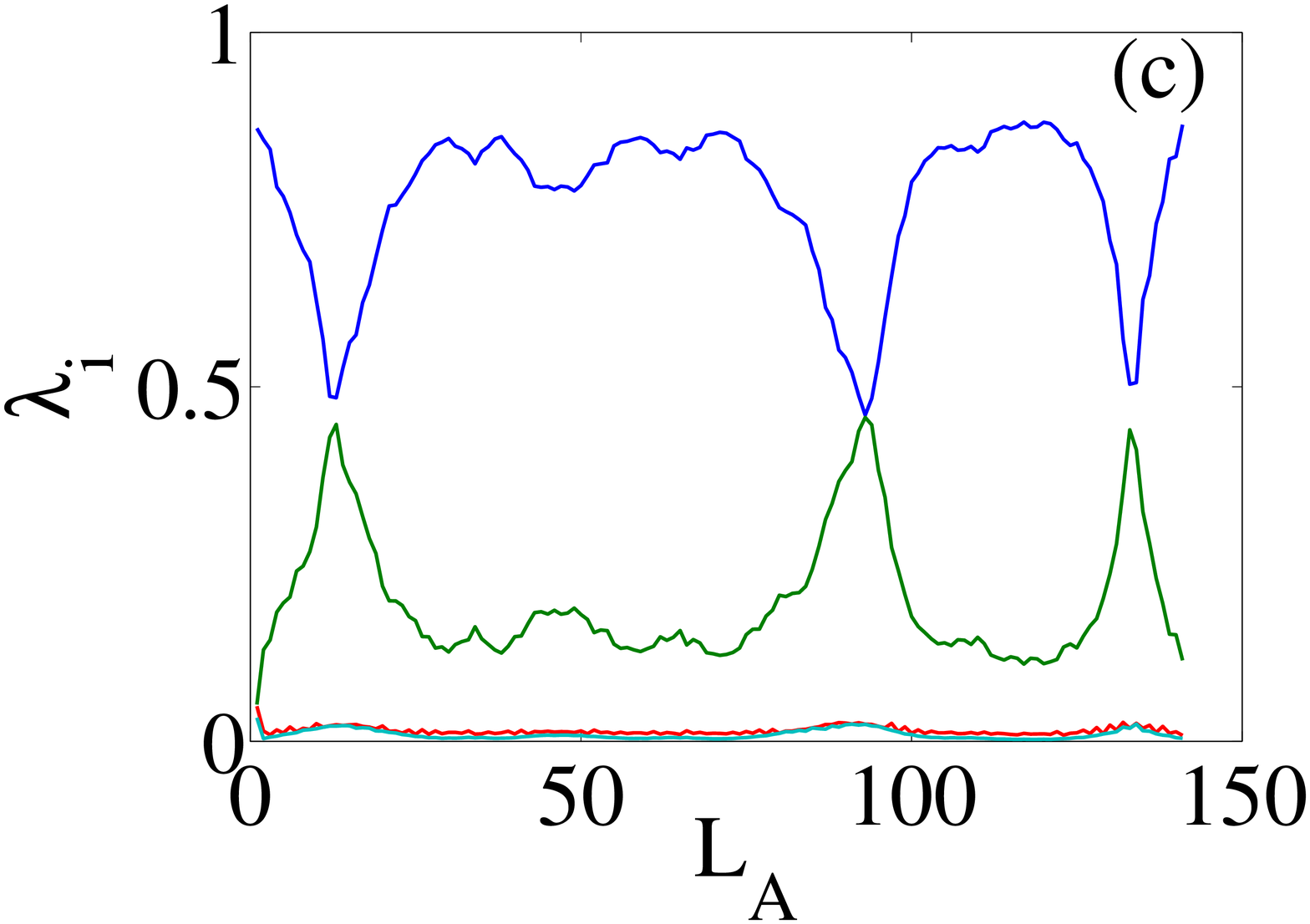}
\includegraphics[width=0.3\linewidth]{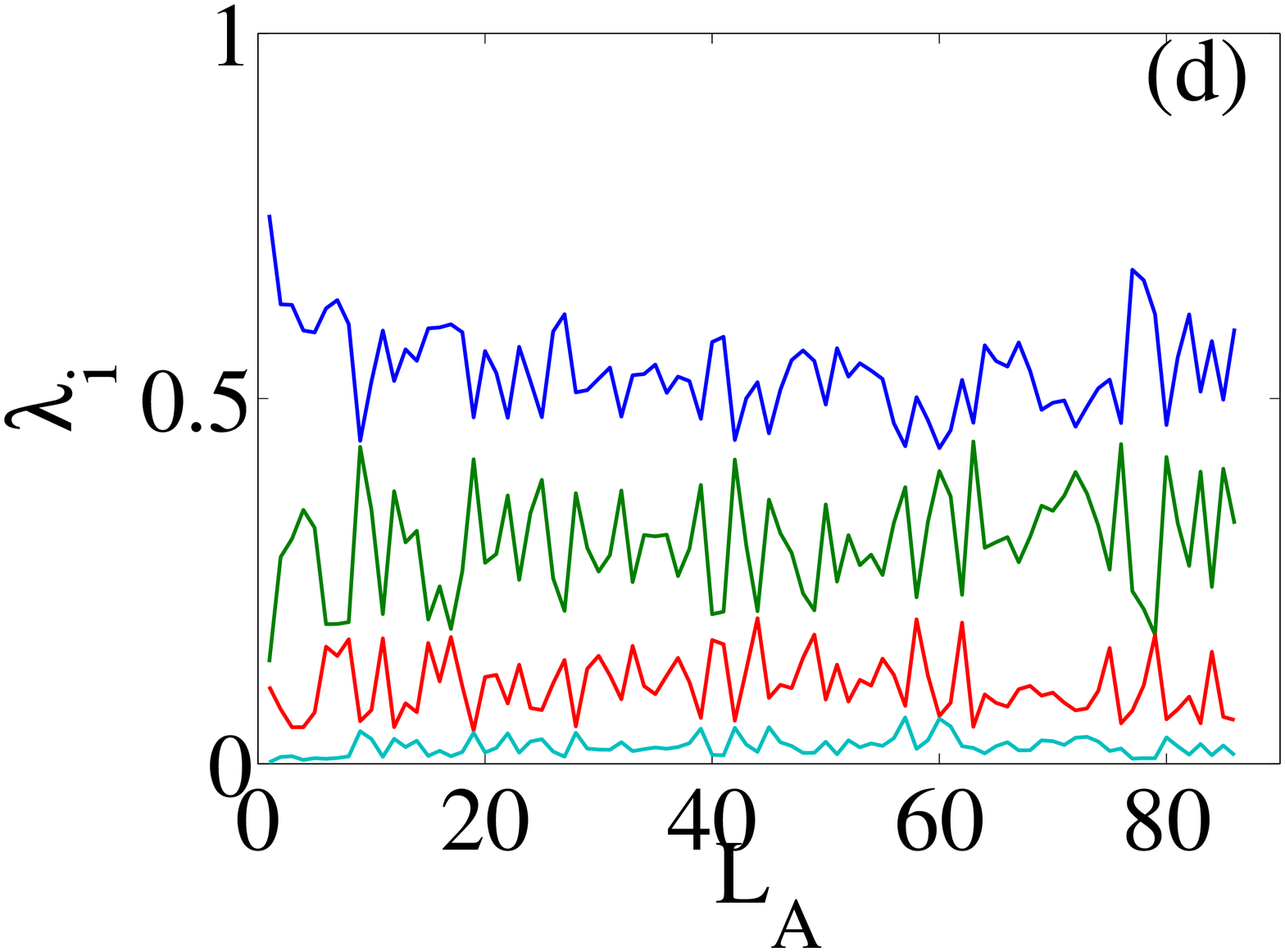}\includegraphics[width=0.3\linewidth]{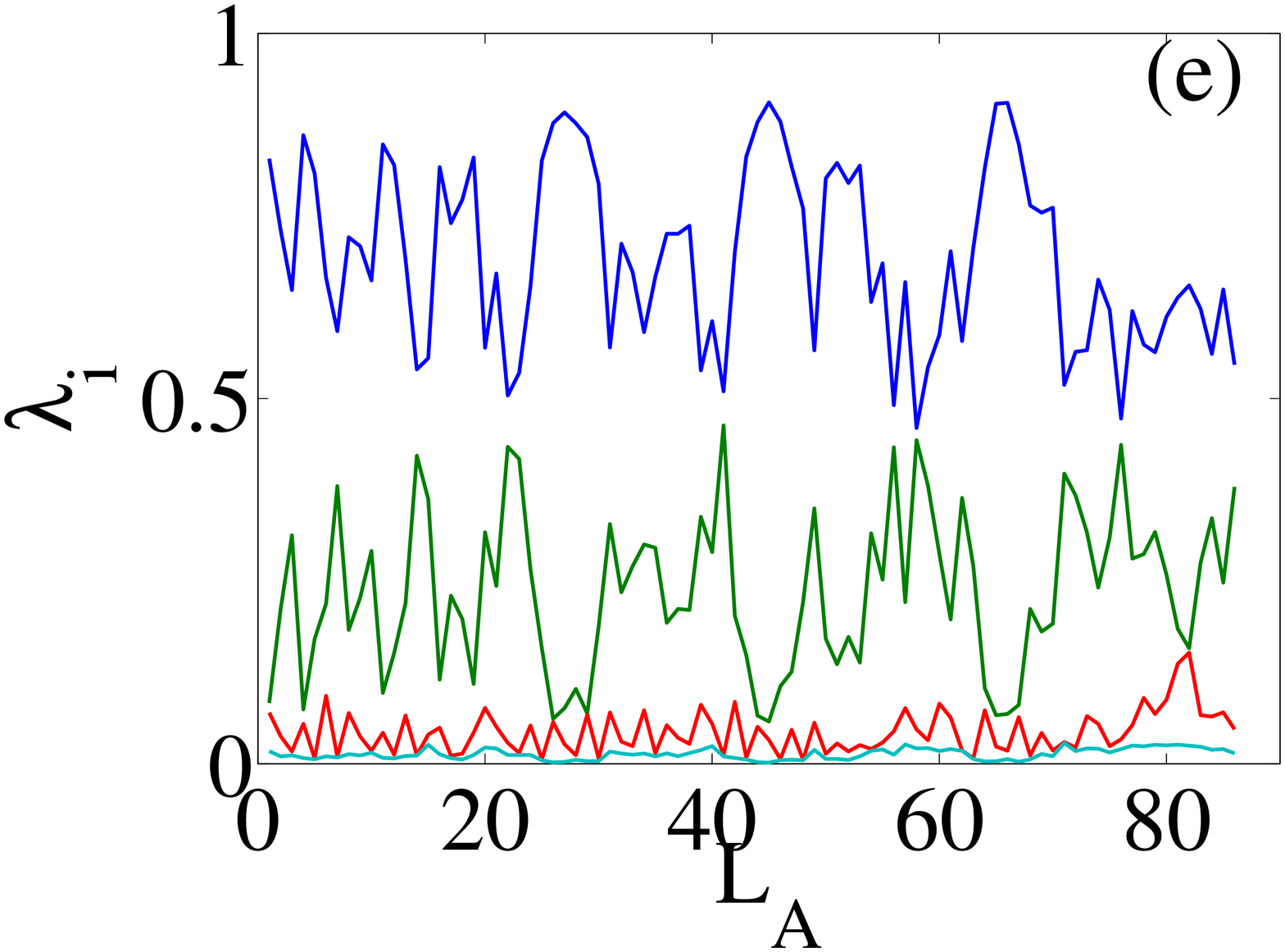}
\end{center}
\caption{Entanglement spectrum of the EBHM in a single realization of the uniform random disorder: largest eigenvalues as a function of the partition along the chain corresponding to a) MI at  $V/t=0.5$ $\Delta/t=0.2$; b) HI at $V/t=3.3$,  $\Delta/t=0.2$; c) BG (remnant of DW) at $V/t=3.6$, $\Delta/t=0.2$; d) SF $V/t=0$, $\Delta/t=3.2$; e) BG at $V/t=3.1$, $\Delta/t=2$. The onsite interaction is $U/t=5$ and $L=89$.}
\label{fig5}
\end{figure}

We consider next the case of a uniform disorder. In Fig.~\ref{fig4} we show the phase diagram obtained from the entanglement spectrum.
Some examples of the largest eigenvalues in the various phases are given in Fig.~\ref{fig5} for a given disorder realization. The main features are the same as in the quasiperiodic case: a unique large eigenvalue for the MI case and a few equally spaced eigenvalues for the SF case, the BG being intermediate between the two above configurations. We notice that  although the DW disappears under finite-size scaling, some remnants of this phase are still visible in the eigenvalues at very low disorder strength.

\section{Conclusions and perspectives}
In conclusion, we have studied the phase diagram of the extended Bose-Hubbard model in the presence of a quasiperiodic potential or a uniform random disorder.  The analysis of the entanglement spectrum is found  to be very helpful in identifying the various phases obtained from the analysis of correlation functions, and well matches with the predictions of a renormalization group approach, when applicable. In perspective, it would be interesting to further explore the nature of the various phase boundaries, extending the work done in the case of the Bose-Hubbard model \cite{Altman,Zoran}.

\section{Acknowledgements}
We thank T. Vekua for useful discussions. X.~D. and L.~S. are supported by the
German Research Foundation~(SA1031/6), the German-Israeli Foundation, and the Cluster of
Excellence QUEST. E.~O. and A.~M. acknowledge support from the CNRS PEPS-PTI project "Strong correlations
and disorder in ultracold quantum gases", and A.~M. from the Handy-Q ERC project.

\end{document}